\documentclass[twoside]{dis04}
\def\runauthor{L.Favart and M.V. Machado}
\def\shorttitle{Deeply Virtual Compton Scattering in the saturation approach}

\newcommand{\rr}{\mbox{\boldmath $r$}}
\newcommand{\rrn}{\mbox{$r$}}

\newcommand{\gsim}{\raisebox{-0.5mm}{$\stackrel{>}{\scriptstyle{\sim}}$}}

\begin{document}
   
\title{Deeply Virtual Compton Scattering in the saturation approach}

\author{Laurent Favart$^{\star}$, M.V.T. Machado
$^{\star\star}$}

\address{\rm $^{\star}$ I.I.H.E., Universit\'e Libre de Bruxelles, Belgium\\
E-mail: lfavart@ulb.ac.be\\
$^{\star\star}$ \rm High Energy Physics Phenomenology Group, GFPAE, IF-UFRGS \\
Caixa Postal 15051, CEP 91501-970, Porto Alegre, RS, Brazil\\
E-mail: magnus@if.ufrgs.br
}

\maketitle

\abstracts{}

\vspace{-1.5cm}

\section{Introduction}

The study of hard exclusive reactions in the Bjorken limit is 
crucial to obtain information on parton dynamical correlations in the
nucleon.
The recent data from DESY $ep$ collider HERA on
exclusive diffractive virtual Compton process \cite{Adloff:2001cn,Chekanov:2003ya,Favart:2003kw} (DVCS)
at large $Q^2$ becomes an important source to study the partons, in
particular gluon,
inside the proton for non-forward kinematics and its relation with
the forward one. 
A considerable interest of the DVCS process comes from the particular
access it gives to these generalized parton distributions (GDP) through
the interference term with the Bethe-Heitler process. 
On the other hand, recently the color
dipole formalism has been provided a simultaneous description of photon
induced process. The inclusive deep inelastic reaction and the photon
diffractive dissociation has been successfully described and the study of
other exclusive process as DVCS is an important test of the color dipole
approach.  The work reported here, summarizing the studies in Refs. \cite{Favart:2003cu,Favart:2004uv}, applies the 
successful saturation model \cite{Golec-Biernat:1998js} to the DVCS
process. The model  interpolates between the small and large dipole configurations and has  its parameters obtained from an adjust to
small $x$ HERA data. Moreover, its QCD evolution has been recently computed \cite{Bartels:2002cj}, which improves the high $Q^2$ data description.

\section{DVCS cross section in the color dipole picture}
 
Based on the color dipole framework, the DVCS process can be seen as 
a succession in time of three factorisable subprocesses: i) 
the photon fluctuates in a quark-antiquark pair, ii) this 
color dipole interacts with the proton target, iii) the quark pair
annihilates in a real photon. As usual, the  kinematic variables are the c.m.s. energy squared $s=W_{\gamma p}^2=(p+q)^2$, 
where $p$ and $q$ are the proton and the photon
momenta respectively, the photon virtuality squared $Q^2=-q^2$ and
the Bjorken scale $x=Q^2/(W_{\gamma p}^2+Q^2)$. The DVCS imaginary part of the amplitude at zero momentum transfer reads as,
\begin{eqnarray}
 {\cal I}m\, {\cal A}\,(s,Q^2,t=0)  = \int_0^1 dz \int d^2\rr \,\Psi_T^*(z,\,\rr,\,Q_1^2=Q^2)\,
\Psi_T(z,\,\rr,\,Q_2^2=0)\,\sigma_{dip}(\tilde{x},\,\rr^2)\nonumber \,,\label{dvcsdip}
\end{eqnarray}
where $\sigma_{dip}(\tilde{x},\,\rr^2)$ is the dipole cross section, which depends on the scaling variable $\tilde{x}=\frac{Q^2+4m_f^2}{(W_{\gamma p}^2+Q^2)}$ and dipole size, $\rr$. The product of the photon wavefunctions (transverse polarization) is given by,
\begin{eqnarray}
\Psi_T^*\,\Psi_T = 
\frac{6\alpha_{\mathrm{em}}}{4\,\pi^2} \, \sum_f e_f^2 \, \left[ f(z)\, \varepsilon_1 \,K_1 (\varepsilon_1 \,\rrn) \,\varepsilon_2
\,K_1 (\varepsilon_2 \,\rrn) + m_f^2 \, \,K_0(\varepsilon_1\,
\rrn)\,K_0(\varepsilon_2\, \rrn)  \right]\,,\nonumber \label{wdvcstrans}
\end{eqnarray}
where $\varepsilon^2_{1,\,2}= z(1-z)\,Q_{1,\,2}^2 + m_f^2$ and $f(z)=[z^2 +
(1-z)^2]$. The quark mass, $m_f$, plays the role of a regulator as $Q^2 \rightarrow 0$. The relative contributions from dipoles of different sizes can be
analyzed with the weight (profile)  function, 
\begin{eqnarray*} 
W(\rr,Q^2) = \int_0^1 \,dz\, \rr\, \Psi_T^*(z,\,\rr,\,Q_1^2=Q^2)\,
\Psi_T(z,\,\rr,\,Q_2^2=0)\,\sigma_{dip}\,
              (\tilde{x},\,\rr^2)\,.
\end{eqnarray*}
 
\begin{figure}[t]
\begin{tabular}{cc}
\psfig{file=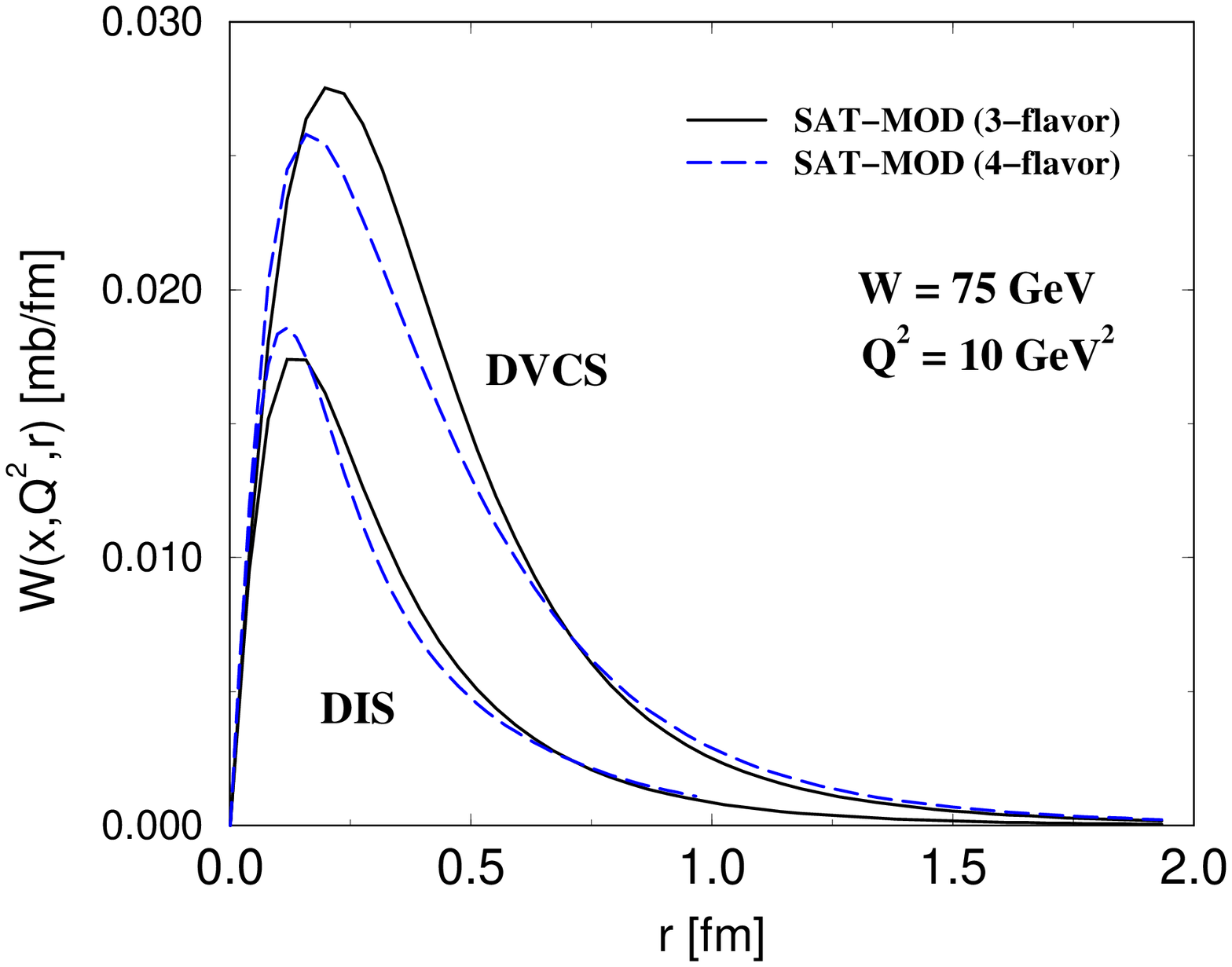,width=0.41\textwidth} & 
\psfig{file=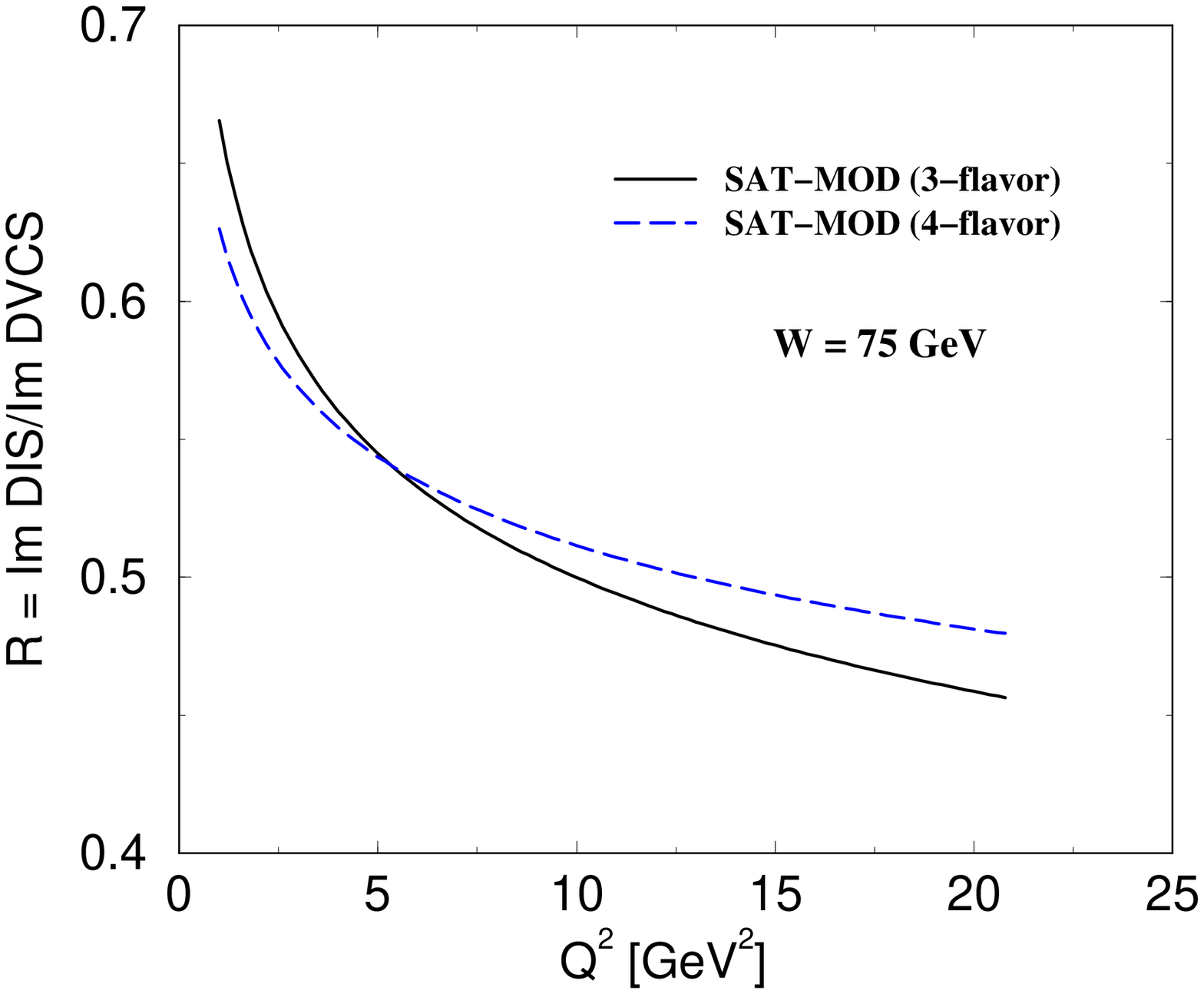,width=0.40\textwidth} \\
(a) & (b)
\end{tabular}
\caption{\it 
 {\bf (a)} Comparison between the profile function $W(\rr,Q^2)$ as a
  function of the dipole size $\rr$ 
  for DVCS and inclusive DIS processes at $Q^2=10$ GeV$^2$.
 {\bf (b)} The ratio of the imaginary parts of the DIS and DVCS
  amplitudes as a function of $Q^2$ at $W=75$ GeV. 
 }
\label{rprofile}
\end{figure}
 
In the DIS process the contribution of large dipole configurations is observed to diminish in a sizable way as the virtualities increase. However, as shown in Fig. (\ref{rprofile}-a), the DVCS profile selects larger dipole sizes in contrast to the
inclusive DIS case, even at relatively large  $Q^2$ (similar result has been  obtained in Ref.\cite{McDermott:2001pt}). The calculation was performed using the saturation model \cite{Golec-Biernat:1998js} for both  three and four-flavor analysis. The inclusion of the charm content gives a lower
normalization for the profile and by consequence for the total cross
section. The impact of the charm is smaller in the inclusive DIS case 
than in DVCS process, confirming that DVCS is more sensitive to the
non-perturbative (soft) content of the scattering process. Moreover, in order to  estimate the importance of the skewing effect,
we calculate the ratio between the
imaginary parts of the forward $t=0$ amplitudes for DIS and DVCS,
$\mathrm{R}={\cal I}m\, \,\mathrm{DIS}/\,{\cal I}m \,\,\mathrm{DVCS}$.
As shown in  Fig. (\ref{rprofile}-b) our result presents values slightly
above those from an aligned jet model analysis
in Ref. \cite{Frankfurt:1997at} and below those from the dipole analysis in
Ref. \cite{McDermott:2001pt}.

The final expression for the DVCS cross section is written as, 
\begin{eqnarray*} 
\sigma(\gamma^*\,p\rightarrow \gamma \,p) =
 \frac{[\,{\cal I}m\,{\cal A}(s,Q^2,t=0)\,]^2}{16\,\pi\,B} \, (1+\rho^2)\,,
\end{eqnarray*}
where $B$ is the $t$ slope parameter and comes from a simple
exponential parameterization and $\rho$, the ratio between the real to
imaginary part, is computed according to the dispersion relations 
(more details in Ref. \cite{Favart:2003cu}).

\section{Cross section comparison to Data}

\begin{figure}[t]
 \begin{center}
  \psfig{file=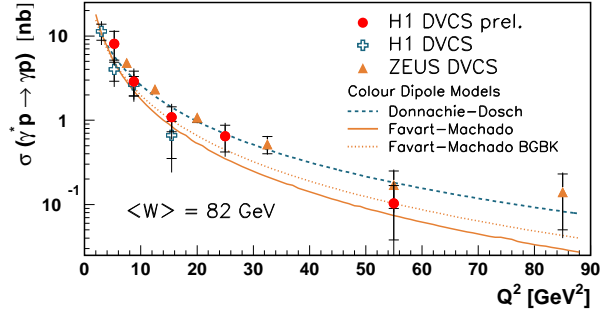,width=0.65\textwidth} 
 \end{center}
 \caption{\it 
  The photon level DVCS cross section as a function of virtuality $Q^2$
  at $W=75$ GeV. Data are compared to the present prediction,
  with and without (BGBK) QCD evolution (see text). }
\label{fig:datacomp}
\end{figure}

In what follows we present the results for the phenomenological saturation models and also its version considering QCD evolution \cite{Bartels:2002cj} (labeled BGBK), giving the dipole cross section gluon dependent.  In Fig. \ref{fig:datacomp} (from the H1 conference paper \cite{Favart:2003kw}) one compares their  prediction with the existing measurement from H1 and ZEUS Collaborations \cite{Adloff:2001cn,Chekanov:2003ya,Favart:2003kw}. The color dipole prediction from Donnachie-Dosh \cite{Donnachie:2000px} is also presented for the sake of comparison.
As the $B$ value has never been measured for DVCS, the normalization of
the theoretical prediction is basically free (usually values of
$5<B<9$ GeV$^{-2}$ are considered, and the fixed value of $B=7$
GeV$^{-2}$ is chosen on the figure). 

Although there is a little difference of normalization between
H1 and ZEUS measurements, which makes difficult to set an overall $B$ 
value for all measurements, the behavior on $Q^2$ and  $W$ (see Ref. \cite{Favart:2003kw}) is well reproduced. On the other hand, for $Q^2\gsim 40 $ GeV$^2$  our prediction still  underestimates the experimental data. 
This change of behavior in the $Q^2$ shape can indicate two situations: (a) the $B$ slope would diminish as increasing virtualities
or; (b) some additional effect should appear at 
higher $Q^2$. 
In order to investigate the first hypothesis, we compute cross section 
using a $Q^2$ dependent slope: $B(Q^2)=8[1-0.15\ln(Q^2/2)]$ GeV$^{-2}$ (see Ref. \cite{Favart:2004uv} for details).
Concerning the second hypothesis, we have investigated two options: QCD evolution (using BGBK model) and skwedness effects. For the skewedness corrections, the ratio of off-forward to forward gluon distribution are  are given explicitly by \cite{Shuvaev:1999ce}, $R_{g}\,(Q^2)=\frac{2^{2\lambda + 3}}
   {\sqrt{\pi}}\,\frac{\Gamma\,\left(\lambda+ 
    \frac{5}{2}\right)}{\Gamma \,\left(\lambda+4 \right)}$,
where $\lambda$ is the effective power on energy of the scattering amplitude. For our purpose the amplitude is multiplied by $R_g$, in order to estimate the size of the skewedness effects.

\begin{figure}[t]
\centerline{\psfig{file=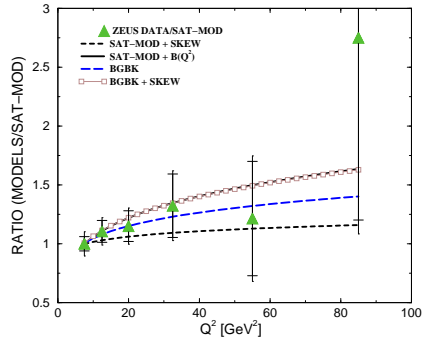,width=55mm}} 
\caption{\it  The ratio MODELS/SAT-MOD as a function of $Q^2$ (see text for details).}
\label{fig:studies} 
\end{figure}

To compare the $Q^2$ dependence, we normalize all models to describe
the ZEUS data point at the lowest $Q^2$ value. 
Further, we plot the ratio of each model to
our baseline model SAT-MOD as a function of $Q^2$. 
Such a procedure allows a $Q^2$ dependence
comparison independently of the normalization effect. These ratios are
shown in Fig.~\ref{fig:studies}, where the points (triangles-up) are the
ratio of the ZEUS data to SAT-MOD including the error bars for the
statistical (inner) and sum in quadrature of statistical and
systematic (outer) uncertainties. 

We verify  that several models can account
for the measured $Q^2$ dependence,  which are not distinguishable
with the present experimental precision.
If the change in normalization is small for the inclusion of a $Q^2$ 
dependence in $B$, the effect is of the order of 12\% for BGBK 
with respect to the basic SAT-MOD and of
40\% for the skewedness effect (SKEW) and still larger when the different
effect are combined (60\% for BGBK+SKEW).  Therefore, these issues show clearly the importance of a measurement of the slope $B$. Such a measurement  would already allow
to discriminate among the different theoretical predictions with
an amount of data comparable to the present ZEUS measurement.


\end{document}